\documentclass[a4paper,fleqn,usenatbib]{mnras}

\usepackage[T1]{fontenc}
\usepackage{ae,aecompl}

\usepackage{graphicx}	% Including figure files
\usepackage{amsmath}	% Advanced maths commands
\usepackage{amssymb}	% Extra maths symbols

\title[Interstellar scintillations of PSR B1919+21]{Interstellar scintillations of PSR B1919+21: space-ground interferometry}

\author[V.I. Shishov et al.]{
V. I. Shishov,$^{1}$\thanks{E-mail: shishov@prao.ru (PRAO ASC LPI)}
T. V. Smirnova,$^{1}$\thanks{E-mail: tania@prao.ru (PRAO ASC LPI)}
 C. R. Gwinn$^{2}$\thanks{E-mail:cgwinn@ucsb.edu (University of California)}
 A.S.Andrianov$^{3}$\thanks{E-mail:andrian@asc.rssi.ru (ASC LPI)}
\newauthor  M. V. Popov,$^{3}$\thanks{E-mail: mwpopov@gmail.com (ASC LPI)}
A. G. Rudnitskiy,$^{3}$\thanks{E-mail: almax1024@gmail.com (ASC LPI)}
V. A. Soglasnov$^{3}$\thanks{E-mail:vsoglasn@asc.rssi.ru (ASC LPI)}
\\
$^{1}$Pushchino Radio Astronomy Observatory, Astro
Space Center, Lebedev Physical Institute, Russian Academy of
Sciences, \\
Pushchino, Moscow oblast', 142290, Russia\\
$^{2}$Department of Physics, University of California,
Santa Barbara, California 93106, USA\\
$^{3}$Astro Space Center, Lebedev Physical Institute, Russian Academy Of Sciences, Profsoyuznaya 84/32, 117997, Russian Federation\\
}

\date{Accepted XXX. Received YYY; in original form 04.05.2016}

\pubyear{2016}

\begin{document}
\label{firstpage}
\pagerange{\pageref{firstpage}--\pageref{lastpage}}
\maketitle

\begin{abstract}
We carried out observations of pulsar PSR B1919+21 at 324 MHz to study the 
distribution of interstellar plasma in the direction of this pulsar. 
We used the RadioAstron (RA) space radiotelescope together with two ground 
telescopes: Westerbork (WB) and Green Bank (GB). The maximum baseline 
projection for the space-ground interferometer was about 60000 km. We show that 
interstellar scintillation of this pulsar consists of two components: 
diffractive scintillations from inhomogeneities in a layer of turbulent 
plasma at a distance $z_{1} = 440$ pc from the observer or homogeneously distributed scattering material to pulsar; and weak 
scintillations from a screen located near the observer at $z_{2} = 0.14 \pm 0.05$ pc. Furthermore, in the direction to the pulsar we detected a prism that deflects radiation, leading to a shift of observed source position. We show that the influence of the ionosphere can be ignored for the space-ground baseline. Analysis of the spatial coherence function for the space-ground baseline (RA-GB) yielded the scattering angle in the observer plane: $\theta_{scat}$ = 0.7 mas. An analysis of the time-frequency correlation function for weak scintillations yielded the angle of refraction in the direction to the pulsar: $\theta_{ref, 0}$ = 110 ms and the distance to the prism $z_{prism} \le 2$ pc.
\end{abstract}


%%%%%%%%%%%%%%%%%%%%%%%%%%%%%%%%%%%%%%%%%%%%%%%%%%

%%%%%%%%%%%%%%%%% BODY OF PAPER %%%%%%%%%%%%%%%%%%

\section{Introduction}

Fluctuations of electron density in the interstellar plasma scatter
radiowaves from astronomical objects. 
The observer at the Earth detects a signal that is a convolution of the initial signal and 
a kernel that describes scattering in the interstellar plasma \citep{GJ}. 
Several effects are observed for pulsars corresponding to the scattering of the radio emission: 
intensity modulation  in frequency and in time (scintillations), pulse broadening, 
angular broadening, and signal dispersion in frequency. 
 
A space radiotelescope such as RadioAstron provides a great opportunity to measure the parameters 
of scattering. Separation of the effects of 
close and distant scattering material requires 
high spatial resolution.
RadioAstron provides the space element of this interferometer for our observations.
Technical and measured parameters of the RadioAstron mission 
have been described by \citet{AA} and \citet{KK}
We observed several close pulsars in the Early Science program of RadioAstron (RAES), including pulsars B0950+08 and B1919+21. First results, published in the paper of \citet{S14} show that 
a layer of plasma located very close to the Earth, at 4.4 to 16.4 pc, is primarily responsible for scintillation of B0950+08. 
First indications that the nearest interstellar medium is responsible for the scintillations of pulsars B0950+08 and J0437-47 were discussed in earlier papers \citet{SS08,SG06}; see also \citet{Bhat2016}. These pulsars have among the lowest dispersion measures observed, indicating a low column density of plasma. 
A scattering medium located at a distance of about 10 pc from the Sun
is also responsible for the variability of some quasars over periods of about an hour, when observed at centimeter wavelengths  \citep{KC, D02, big03, den03, Jau03, B06}. These observations of scattering of close pulsars and short-period variability of quasars indicate the existence of a nearby interstellar plasma component
that has properties different from those of more distant plasma components.  
 
The aim of the study reported here is to investigate the spatial distribution 
of the interstellar plasma toward the pulsar B1919+21. We show 
that two isolated layers of 
interstellar plasma lie in this direction, one of which is localized at a distance of only 
 0.14 pc. Pulsar B1919 + 21 is a strong pulsar. 
Its period is $P_1=1.3373\ {\rm s}$.
It lies at galactic 
latitude $3.5^{\circ}$ and longitude $55.8^{\circ}$. Its 
dispersion measure is ${\rm DM} = 12.43\ {\rm pc\ cm}^{-3}$. The
\citet{CL} model  indicates that the pulsar distance is 1 kpc. Measurements of this pulsar's proper motion yielded
$\mu_{\alpha} = 17 \pm 4\ {\rm mas/yr}$ and 
$\mu_{\delta} = 32 \pm 6\ {\rm mas/yr}$ \citep{Z05}. 
 
\section{Observations}

%Normally the next section describes the techniques the authors used.
%It is frequently split into subsections, such as Section~\ref{sec:observations} below.

We conducted observations of PSR B1919+21 at an observing frequency of 324 MHz 
on 4 July 2012, using the RadioAstron 10-m space 
radiotelescope together with 110-m Green Bank (GBT) and $14 \times 25$-m Westerbork 
(WSRT) telescopes. Data were transferred from RadioAstron in  
real time to Puschino, where they were recorded using the RadioAstron 
Digital Recorder (RDR) \citep{AG},
developed at the Astro-Space Center of the Lebedeev Physical Institute (ASC). The Mark5B recording system was used for the ground 
telescopes. All telescopes recorded the frequency band from 316 to 332 MHz, with one-bit 
quantization for space telescope data, and two-bit quantization for 
ground telescopes. 
Data were recorded 
for 4170 s, divided into scans of $421\  P_1$ (563 s) and subintervals of $26\ P_1$ (about 35 s). 
The primary data processing was done using the ASC 
correlator \citep{AG} with incoherent dedispersion. 
Data were correlated with 
512 spectral channels in two selected windows: on pulse and off 
pulse, the width of each window was 40 ms (3\% of the pulsar period). An on-pulse window was centered on the maximum of 
the average profile, and an off-pulse window was selected at 
half the pulsar period from the on-pulse window. The projected space-ground interferometer baseline 
was about 60,000\ km. 

\section{Data processing and analysis}

\subsection{Dynamic Spectrum and Correlation Functions}

\subsubsection{Dynamic Spectrum}

\begin{figure*}
\includegraphics[scale=0.5]{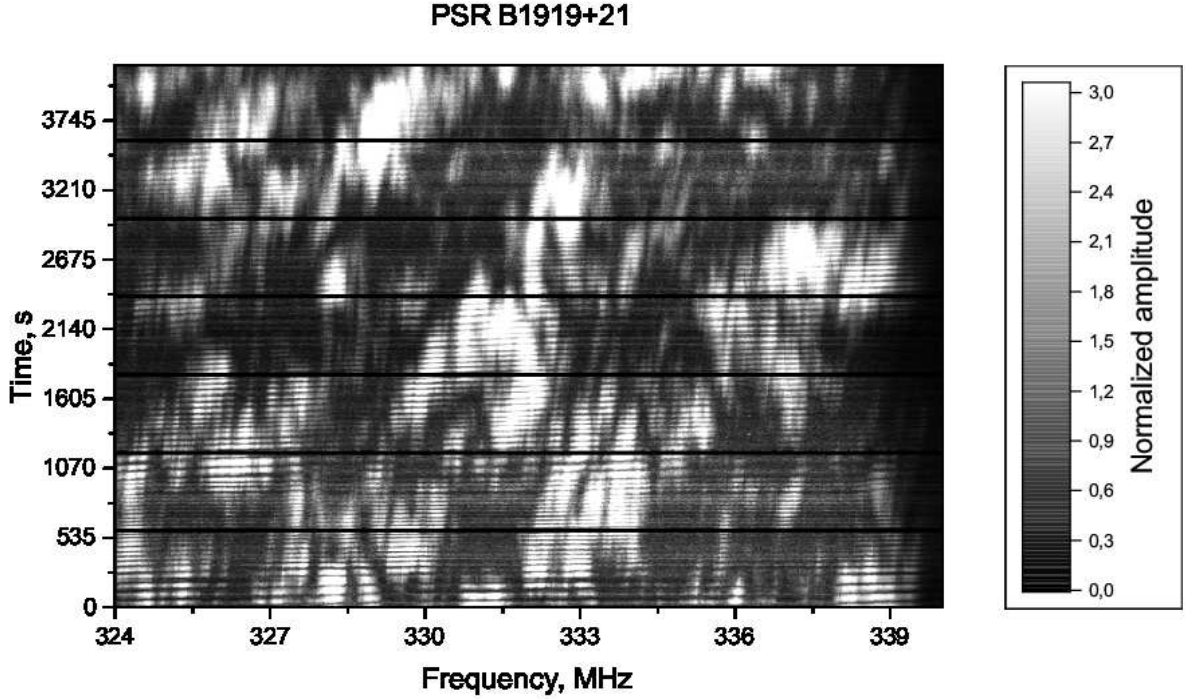}
\caption{Dynamic spectrum of PSR B1919+21 at observing frequency 324 MHz, for 
the ground baseline Green Bank - Westerbork (GB-WB).
 Grayscale shows normalized amplitude.}
\label{fig:fig1}
\end{figure*}

We formed complex cross-spectra 
between pairs of telescopes for all scans, in on- and off-pulse windows. 
In some cases, to 
increase the sensitivity, we averaged cross-spectra over 
4\ pulsar periods. To obtain dynamic spectra, we calculated the modulus of the % mmodulus OK
cross-spectra, and corrected for the receiver bandpass using the 
off-pulse spectra. To reduce the impact of broadband intensity 
variations of the pulsar from pulse to pulse, we normalized each spectrum by its standard deviation, 
$\sigma(t)$. Figure\ \ref{fig:fig1} shows the normalized dynamic spectrum of scintillation of pulsar B1919+21 
for the Green Bank - Westerbork ground interferometer
(GB-WB). We see clearly expressed large-scale sloping 
structures (slanting features), with scales of $df=1$\ to\ 1.5\ MHz in 
frequency, and of $d t\approx 1000$\ s in time. Diffractive spots are strongly extended along the line $f = (df/dt) t$. This drift indicates 
that refraction by a cosmic prism determines the structure of scintillation in the frequency-time domain. 
The regular dark bands along the frequency axis represent the intervals when signal was not recorded,
and were 
filled with values of zero.
The narrower gray bands show the pulse-to-pulse variability intrinsic to the pulsar. 

\subsubsection{Drift and Dual Frequency Scales}\label{sec:DriftAnd2Scales}

\begin{figure}
\includegraphics[width=\columnwidth]{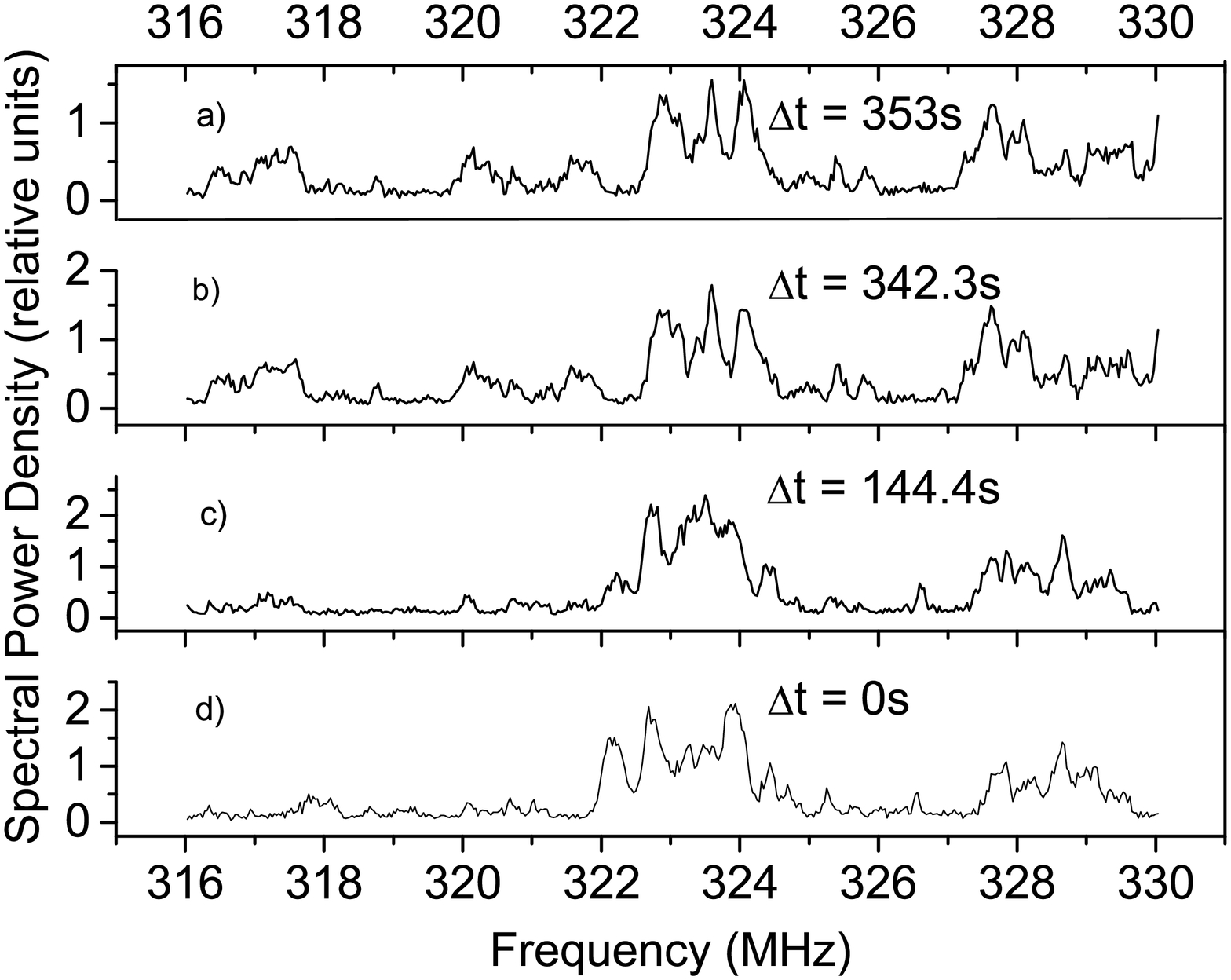}
\caption{Spectra of individual pulses of 
the pulsar, separated in time by the specified number of seconds from the 
bottom spectrum.}
\label{fig:fig2}
\end{figure}

To determine the drift rate of the 
diffractive structure, we calculated the position of maximum of  the mean cross-correlation
between spectra separated in time by lags $k P_1$, where $k = 1, 2, \ldots$ and $P_1$ is a pulse period in s.  
We found three separated
 bands with the slope $\Delta f/\Delta t = 1.5$\ MHz/1000s.  We obtained this slope by a least-squares fit  to the 
positions of maximums.
Figure \ref{fig:fig2} shows spectra of several strong pulses 
separated in time, with time increasing from bottom to top in the figure. 

Two scales of structure are visible in the  
spectra: small-scale structure with a frequency scale of about 400 kHz, and 
large-scale structure with frequency scale of about 1500 kHz. 
These scales are the approximate full width at half-maximum amplitude of the features.
At smaller separations in time (as in spectra a and b in the figure, separated by 11\ s), the fine 
structure is the same. Over longer separations (as in spectra b and c, separated by 200\ s) the fine structure changes, but
the large-scale structure retains its shape. The modulation 
index, defined as $m(t) = \sigma(t)/\langle I\rangle_f(t)$, 
varies from 0.7 to 1.0 on a time scale of the order of 500 s. This 
variation indicates that the statistics are not sufficient to 
determine the correct value of $m(t)$. 
However, the fact that the modulation index is close to 1 confirms that the scattering is strong.

\subsubsection{Determination of Frequency Scales}\label{sec:DeterminationOfFreqScales}

\begin{figure}
\includegraphics[width=\columnwidth]{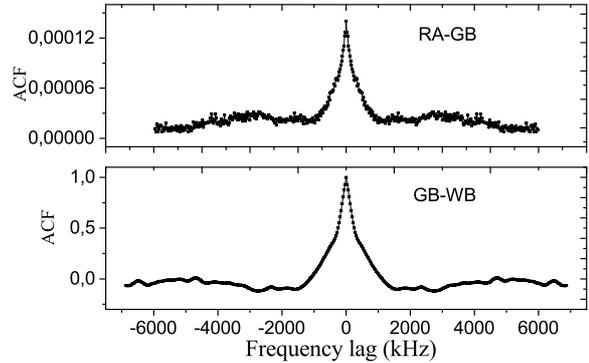}
\caption{Autocorrelation functions (ACF) 
averaged over the entire observation time for space-ground (RA-GB, upper) and ground (GB-WB, 
lower) baselines. Expression\ \ref{eq:10}
was used to calculate the ACF for the space-ground baseline.}
\label{fig:fig3}
\end{figure}

A correlation analysis of the dynamic spectra provides the scales of scintillation in frequency, $\Delta f_{dif}$. Figure \ref{fig:fig3} 
shows the average autocorrelation functions (ACF) as a function of  
frequency lag. 
The ACF was averaged over the entire observation.
For our
ground baseline we calculated the ACF using the usual procedure, because the influence of noise 
was small, and because it was necessary to eliminate the 
influence of the ionosphere, as discussed below. 
For the space-ground 
baseline, we calculated the ACF as the modulus of the % modulus tania specifically corrected -- thanks -- checked this again CRG
average correlation function of the complex cross-spectra. The 
corresponding expressions are presented in the Section\ \ref{sec:theor_relations}. 
This 
procedure is required when the contribution of the noise is 
greater than or comparable to the  signal level, and when ionospheric effects are small, as was the case for 
our space-ground baseline. Otherwise, the contribution of noise will distort 
the ACF.

The visible break in the slope of the ACF near a lag of $\pm 300$\ kHz for the ground baseline (GB-WB: Figure \ref{fig:fig3}, lower), 
indicates the presence of structure on two scales. 
No break appears in the ACF for the space-Earth baseline:
its shape corresponds to the small-scale structure only. 
To determine the widths and the relative amplitudes of these 
structures we fit the ground baseline with  a sum of exponential and Gaussian 
functions. We obtained frequency scales of $\Delta f_{dif}$ = 330 kHz and 
$\Delta f_{wide}$ = 700 kHz (half width at half-maximum amplitude), with amplitudes of 0.84 and 0.15 for the small- and large-scale structures, 
respectively. As it will be shown below in Section 4, the small-scale structures arise 
from scattering of radiation in the distant layer 
of the turbulent medium as diffractive scintillation, and the large-scale structures in the layer located close to observer as weak 
scintillation.

\subsubsection{Determination of Time Scale}\label{sec:DeterminationOfScales}

\begin{figure}
\includegraphics[width=\columnwidth]{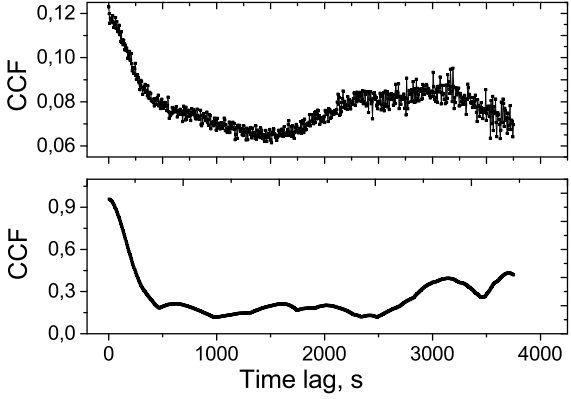}
\caption{The dependence of the average 
cross-correlation coefficient between pairs of spectra spaced 
by the corresponding time interval $\Delta t= 4 P_1 k,\; k = 1, 2, ..$, for 
space-ground (upper) and ground (lower) baselines.}
\label{fig:fig4}
\end{figure}

Figure \ref{fig:fig4} shows the average cross-correlation 
coefficient between pairs of spectra as a function of time
for space-ground (upper) and ground (lower) 
baselines.
The time separations are $\Delta t = 4 P_1 k$, where $k = 1, 2,...$. 
Both baselines show 
the same scintillation scale: 
$\Delta t_{dif}$ = 290 s, expressed as the time lag of $1/e$ of the peak amplitude.

\subsection{Theoretical relations, ionospheric effects, correlation functions}
\label{sec:theor_relations} % used for referring to this section from elsewhere

Much of our analysis in this paper follows that of \citet{S14} in general outline, but here we describe 
new details connected with ionospheric effects and use different definitions for the correlation functions.
Let $h ( f, t)$ be the spectrum of the field of initial pulsar emission 
in the absence of any turbulent medium, where $ f  =  \nu - \nu_0$   is the
offset of the observing frequency $\nu$ from the band center $\nu_0 = 324$\ MHz, and $t$  is
time.
This spectrum
$h ( f, t)$ also includes instrumental modulation of emission in 
the passbands of the receivers. After propagation through the  
turbulent interstellar medium, the spectrum of the electric field for one 
antenna can be represented as: 
\begin{equation} 
E (\vec \rho,  f, t) =  h ( f, t) u (\vec \rho,  f, t) 
\exp\left[-i S (\vec \rho,  f,t)\right], 
\label{eq:1} 
\end{equation} 
where the modulation coefficient $u (\vec \rho,  f, t)$ is determined by propagation through the  interstellar medium, and $\vec \rho$ is the spatial coordinate in the 
observer plane, perpendicular to the line of sight. 
The phase $S(\vec \rho,  f, t)$ is determined by the 
ionosphere and cosmic prism. Multiplying $E (\vec \rho,  f, t)$ 
by $E^* (\vec \rho + \vec b,  f, t)$ and averaging over the 
statistics of the source, we obtain the quasi-instantaneous response of an interferometer with a baseline $\vec b$, the cross-spectrum of the electric field: 
\begin{align} 
I(\vec \rho, \vec \rho +\vec b,  f, t) & = E(\vec \rho,  f, t) 
E^* (\vec \rho + \vec b,  f, t)  \label{eq:2} \\
& = H( f, t) j (\vec \rho, \vec \rho + \vec b,  f, t)   \nonumber \\ &  \quad \times   \exp[-i 
\Delta S (\vec \rho, \vec b,  f, t)] \nonumber 
\end{align}
where:
\begin{align}
j(\vec \rho, \vec \rho + \vec b,  f, t) &= 
u (\vec \rho,  f, t) u^* (\vec \rho + \vec b,  f, t) \label{eq:3} \\
H ( f, t) &= \langle h ( f, t) h^* ( f, t) \rangle_h  \label{eq:4} 
\end{align} 
The subscript $h$ indicates 
averaging over the statistics of the noiselike electric field of the source. 
We assume that the intrinsic spectrum of the source, and of our instrumental response, is flat: $ H( f,t) = 1$. 
The phase difference between the antennas at either end of the baseline
$\Delta S (\vec \rho,  f, t)$ consists of two components, from interstellar refraction and from the ionosphere: 
\begin{equation} 
\Delta S(\vec \rho, \vec b, f,t)  =  \Delta S_{ion} (\vec \rho, 
\vec b, f, t) + \Delta S_{ref} (\vec b, f) \label{eq:5} 
\end{equation} 
For a fixed baseline,
the refractive component of the interferometer phase, $\Delta S_{ref} 
(\vec b,  f)$ depends only on $ f$: 
\begin{equation} 
\Delta S_{ref} (\vec b, f)  =  2\pi \left({\frac{ f}{c}}\right) \, \vec b \cdot \vec 
\Theta_{ref,0}
\label{eq:6}
\end{equation} 
Here $\vec \Theta_{ref,0}$ is the refraction angle at frequency 
$\nu_0$. 
 
The ionospheric component can be represented as 
\begin{align} 
\Delta S_{ion} (\vec \rho,\vec b,  f,t) & =  \Delta S_{ion}(\rho,\vec b,
 f = 0,t) +
\frac{ f}{\nu_0} \Delta S_{ion,0}    \nonumber \\ & \quad +  \frac{ f}{\nu_0} \frac{(t-t_0)}{T} 
\Delta S_{ion,1} \label{eq:8} \\ 
\Delta S_{ion,0} & =  \Delta S_{ion} (\vec \rho,\vec b, f = 0,t = t_0) \nonumber\\ 
\Delta S_{ion,1} & =  T \frac{d}{dt}[\Delta S_{ion} (\vec \rho,
\vec b, f = 0,t)] \mid_{t = t_0}  \nonumber
\end{align} 
where $T$ is the time span of the observations, and $t_0$ is the time at the middle. 
  
\begin{figure}
\includegraphics[width=\columnwidth]{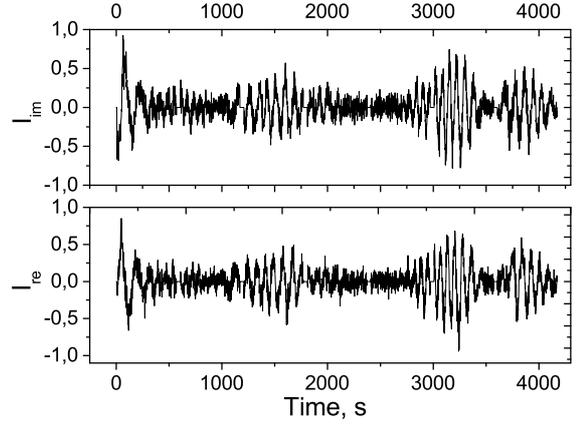}
\caption{Dependencies of the real (upper) and 
imaginary (lower) parts of the complex cross-spectrum on time 
for frequency channel 100, for the Green Bank-Westerbork
baseline (GB-WB). Data were not recorded during intervals indicated as zero amplitude.
\label{fig:fig5}}
\end{figure}

Figure\ \ref{fig:fig5} shows the values of the real and imaginary parts of the interferometer response for one selected 
frequency channel as a function of time:
${\rm Re}[I (\rho, \rho + b,  f, 
t)]$ (upper) and ${\rm Im} [I (\rho, \rho + b,  f, t) $] 
(lower). In addition to the 
amplitude fluctuations corresponding to the dynamic spectrum, 
we see periodic fluctuations with a characteristic period of 
about 70 s, phase-shifted by $90^{\circ}$. Changes of the ionosphere in time cause these fluctuations. Therefore, to 
analyze the data from the ground interferometer we must work 
with the moduli of the cross-spectra. For the space-ground interferometer, % mmodulus -- inferred from tania comments
the influence of ionosphere was much 
smaller, as will be shown below, and so data processing used the complex 
cross-spectra. 

Multiplying $I (\vec \rho, \vec \rho + \vec b,  f, t)$  by its complex conjugate  at frequency 
$ f + \Delta f$  (where $\Delta f =$ frequency shift) and averaging over time and frequency, we obtain 
\begin{align} 
&\langle I (\vec \rho, \vec \rho + \vec b,  f, t) I^* (\vec \rho, \vec \rho + \vec b,  f + \Delta f, t) \rangle  
\label{eq:9} \\ 
&\quad =\langle j (\vec \rho, \vec \rho + \vec b,  f, t)  j^* (\vec \rho, 
\vec \rho + \vec b,  f + \Delta f, t) \rangle \, \varphi (\Delta f)             \nonumber \\ 
&\quad  \times            \exp[ - i (\Delta  f / \nu_0) 
[\Delta S_{ion,0}  +  \Delta S_{ref,0}] 
\nonumber
\end{align} 
where:
\begin{equation} 
\begin{split} 
\varphi (\Delta f) = \left\langle \exp\left[-i \left( \frac{\Delta f }{ \nu_0 }\right) \frac{(t - t_0) }{ T} \Delta 
S_{ion,1}\right] \right\rangle_t  \\ = 
 \frac{ \sin\left[(\Delta f / 2 \nu_0 ) \Delta S_{ion,1} \right]}{(\Delta f / 
2 \nu_0 ) \Delta S_{ion,1}} \label{eq:11} 
\end{split} 
\end{equation} 
In Equation\ \ref{eq:11}, the average corresponds to an integration over $t$ from 
$(t_0  - T/2)$ to $(t_0  + T/2)$. The phase difference at 
frequency $\nu_0$ is $\Delta S_{ref,0}$.
 
The modulus of the averaged correlation in frequency $f$ of the interferometer response $I$ is:  % mmodulus OK
\begin{equation} 
J_1(\vec b, \Delta f) = | \langle I (\vec \rho, \vec \rho + \vec b, 
 f, t) I^* (\vec \rho, \vec \rho + \vec b,  f + \Delta f, t) 
\rangle|  \label{eq:10} 
\end{equation} 
In contrast, the averaged modulus of the correlation in frequency of $I$ is:   % mmodulus OK
\begin{align} 
J_2 (\vec b, \Delta f)  &= \langle | I (\vec \rho, \vec \rho + \vec 
b,  f, t) I^*(\vec \rho, \vec \rho + \vec b,  f + \Delta f, 
t) | \rangle \label{eq:13}  \\
& =  \langle j (\vec \rho, \vec \rho +\vec b, 
 f, t) j^* (\vec \rho, \vec \rho + \vec b,  f + \Delta f, t) 
\rangle \nonumber
\end{align}
 
Figure \ref{fig:fig3} (upper panel) shows the modulus of the averaged correlation function of $I$ for the space-ground interferometer,   % mmodulus OK
as defined in Equation\ \ref{eq:10}. 
The imaginary part of the second moment divided by its modulus is $\varphi_1 (\Delta f)$. The function $\varphi_1 (\Delta f)$ is proportional to $\varphi(\Delta f)*\sin[-\Delta f/\nu_0(\Delta S_{ion,0}  
+  \Delta S_{ref,0})]$. 
 
\begin{figure}
\includegraphics[width=\columnwidth]{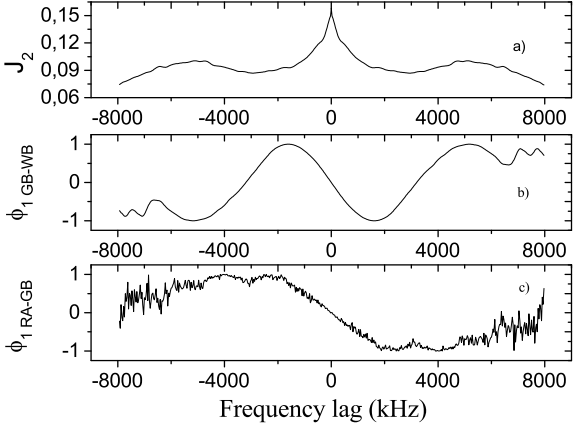}
\caption{a) Average modulus of the correlation of $I$ with frequency offset $f$ (Equation\ \ref{eq:13}; % mmodulus OK
b) $\varphi_1 (f)$, the ratio of the imaginary part of the 
correlation to its modulus for the ground interferometer (GB-WB) % mmodulus OK
as a function of $f$; c) the same ratio $\varphi_1 (f)$ for the space-ground % mmodulus OK
interferometer (RA-GB).}
\label{fig:fig6}
\end{figure}

Figure \ref{fig:fig6} shows values of $J_2(\Delta f)$ and $\varphi_1 (\Delta f)$ for the ground interferometer,
and $\varphi_1 (\Delta f)$ for the 
space-ground interferometer. We see that for the ground 
interferometer the ionospheric phase is larger and varies more rapidly with frequency than 
for the space-ground interferometer. 

The value of $\varphi_1 (\Delta f)$ is zero at $\Delta f = 0$, as
defined by the factor $\sin[-\Delta f/\nu_0(\Delta S_{ion,0}  +  \Delta S_{ref,0})]$. 
However, the zeros of $\varphi_1 (\Delta f)$  
for the ground interferometer at $\Delta f = \pm 3\ {\rm MHz}$ are defined by $\varphi(\Delta f)$. 
If we set
the argument of $\varphi$ in Equation\ \ref{eq:11} equal to $\pi$  at $\Delta f=3\ {\rm MHz}$, we obtain $\Delta S_{ion,1} \approx 600\ {\rm radians}$ for
the ground interferometer, and $\Delta S_{ion,1} \le  200\ {\rm radians}$ for the space-ground interferometer. From these values, it follows that the response of the ground 
interferometer is greatly distorted by the ionospheric phase, but the 
signal amplitude greatly exceeds the noise. Thus, we use Equation\ \ref{eq:13}
to
calculate the correlation function of amplitude fluctuations for the ground baseline. For the space-ground interferometer, the situation 
is reversed: noise exceeds the signal, and we cannot use Equation\ \ref{eq:13}. However, we can neglect the phase 
distortion of the interferometer response, and use Equation\ \ref{eq:10} to determine correlation 
functions of amplitude fluctuations on 
the space-ground baseline. 

\subsection{Structure functions of the interferometer response fluctuations} 

Structure functions provide insight into the correlation function,
as discussed in earlier papers \citep{S14}.
For the space-ground interferometer on baseline $\vec b_s$ we calculate the structure function in frequency difference $\Delta f$ and time difference $\Delta t$ using 
$J_1(\vec b_s,\Delta f = 0, \Delta t=0) - J_1(\vec b_s,\Delta f, \Delta t)$. We normalize
this expression by $J_1(\vec b_s,\Delta f =0, \Delta t = 0) - J_1(\vec b_s,\Delta f = \Delta f^*,\Delta t = 0)$, where $\Delta f^* \gg \Delta f_{dif}$. Hence we obtain the structure function for the space-ground interferometer:
\begin{eqnarray} 
%\begin{split}
&&SF_s(\vec b_s, \Delta f, \Delta t) 
\\
&&\quad = \frac{\left(J_1 (\vec b_s,\Delta f = 0, \Delta t = 0) - J_1 (\vec b_s, \Delta  f, \Delta t) \right)}
{\left (J_1(\vec b_s, \Delta f =0, \Delta t=0) - J_1(\vec  b_s,\Delta f^*,\Delta t=0 ) \right)}               
%\end{split}
\nonumber
\end{eqnarray} 
Similarly, the structure function for the ground interferometer with baseline $\vec b_g$ yields the normalized structure function 
\begin{eqnarray} 
% \begin{align} 
&&SF_g( \vec b_g,\Delta f, \Delta t)\label{eq:SF_def} 
\\ 
&&\quad = \frac{\left(J_2 (\vec b_g,\Delta f = 0, \Delta t = 0) - J_2 (\vec b_g, \Delta f, \Delta t) \right)}
{\left(J_2(\vec  b_g, \Delta f =0, \Delta t=0) - J_2(\vec  b_g, \Delta f^*,\Delta t=0)\right)} 
\nonumber
\end{eqnarray} 
% \end{align}

\begin{figure}
\includegraphics[width=\columnwidth]{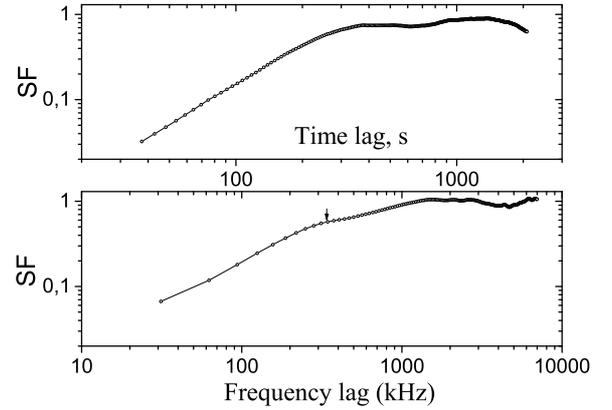}
\caption{Average time (upper) and frequency 
(lower) structure functions of intensity variations (SF) for Green Bank - Westerbork 
(GB-WB) baseline, presented on a log-log scale. The arrow marks 
the frequency lag of the observed break in the structure 
function,  300 kHz.}
\label{fig:GBWB_SF}
\end{figure}

Figure\ \ref{fig:GBWB_SF} shows average time (upper) and frequency (lower) structure 
functions (SF) of intensity variations for our ground baseline on log-log axes. 

An arrow marks the break in the structure function at a frequency lag of $300\ {\rm kHz}$.
We fit power-laws to the structure functions:
\begin{align}
SF(\Delta f) &= \textstyle{ \frac{1}{2} } \left({\Delta f}/{\Delta f_{dif}} \right)^{\beta_f} \\
SF(\Delta t) &= \textstyle{ \frac{1}{2} } \left({\Delta t}/{\Delta t_{dif}}\right)^{\beta_t} \nonumber
\end{align}
We performed linear fits to the logarithmic data for these two structure functions over the ranges
$\Delta t_{samp}<\Delta t <\Delta t_{dif}$ and $\Delta f_{samp}<\Delta f <\Delta f_{dif}$
respectively, where $\Delta t_{samp}$ and $\Delta f_{samp}$ are  the sampling intervals in time and frequency for our data. 
We obtained 
$\beta_f= 0.90\pm 0.03$ for the frequency structure function and $\beta_t = 
1.73 \pm 0.02$ for the time structure function. 
The resulting relation between frequency and time structure functions $\beta_f=\beta_t/2$ 
corresponds to a diffractive model for
scintillation \citep{Sh03}. The power-law index $n$ of the spectrum of density inhomogeneities responsible for scattering
is connected with the index of $SF(t)$ through the relation: $n =\beta_t + 2 = 3.73$   \citep{Sh03}.

\begin{figure}
\includegraphics[width=\columnwidth]{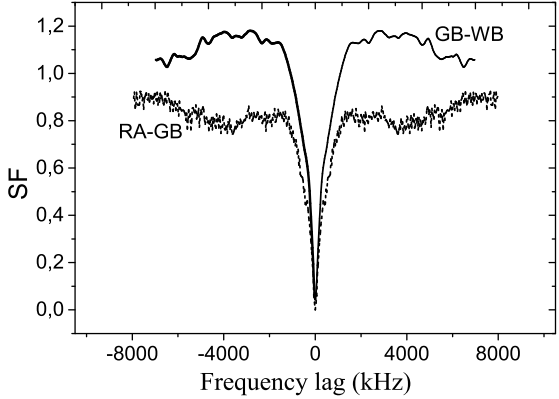}
\caption{Normalized frequency structure 
functions  of intensity variations  for ground (GB-WB) and space-ground (RA-GB) baselines at zero time shift.}
\label{fig:ground_space_freq_0t_SF}
\end{figure}

Figure
\ref{fig:ground_space_freq_0t_SF} shows the average frequency structure functions for the 
ground (line) and space-ground (dash line) baselines at zero 
time shift. Evidently the levels of the SF differ by about 0.2 - 0.3.  
This corresponds to the relative contributions of the two 
frequency scales in the scintillation spectra to the ground baseline, as seen in Figures\ \ref{fig:fig2} and\ \ref{fig:fig3},
and discussed in Section\ \ref{sec:DeterminationOfFreqScales} above. The ratio of their amplitudes is consistent with our fit to two components in the average frequency correlation function 
for the ground baseline. The space-ground baseline shows no such
break in the structure function in Figure\ \ref{fig:ground_space_freq_0t_SF},
or what would be corresponding structure in the correlation function shown in Figure\ \ref{fig:fig3}.
Rather, 
the space-ground structure function displays only the narrower frequency-scale component.  

\begin{figure}
\includegraphics[width=\columnwidth]{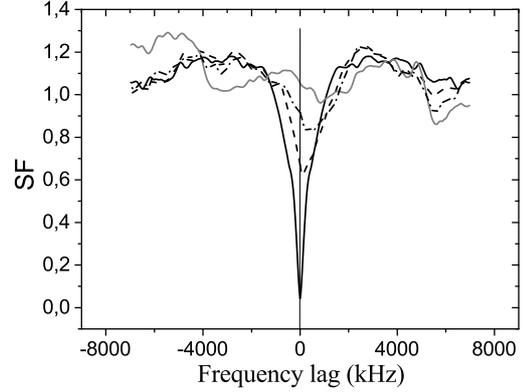}
\caption{Average structure functions  of intensity variations  for 
Green Bank - Westerbork baseline, calculated from spectra for 
different time shifts: black line for $\Delta t$ = 4$P_1$; dash line  
for $\Delta t$ = 200$P_1$; dash-dot line for $\Delta t$ = 320$P_1$; 
grey line for $\Delta t$ = 640$P_1$}
\label{fig:ground_space_freq_deltat_SF}
\end{figure}

Figure
\ref{fig:ground_space_freq_deltat_SF} shows the 
mean frequency structure functions for the ground baseline 
at different time lags: 
4$P_1$ (squares), 200 $P_1$ (circles), 320 $P_1$ (line) and 640 
$P_1$ (triangles). With increasing time shift between 
spectra, the amplitude of the structure function decreases, and its minimum is displaced. 
When $\Delta t = 200 
P_1$ (267 s) the structure function still shows a small contribution of small-scale 
structure, whereas at $\Delta t = 640 P_1$ (856 s) we see only one 
component, the center of which is shifted to 1100 kHz, with amplitude of 0.15. 
Weak scintillation alone would produce a wide-bandwidth pattern; but
the cosmic prism slants the pattern in both frequency,
by dispersion; and in time, by a spatial displacement that the motion of the source converts to the time domain.
Thus, a frequency shift compensates for the time offset, as Figure\ \ref{fig:ground_space_freq_deltat_SF} displays.
This effect is also 
clearly visible in the dynamic spectrum 
(Figure \ref{fig:fig1}) as was discussed in Section\ \ref{sec:DriftAnd2Scales}. The drift in frequency at the rate $d f/d t = 1.1$\ MHz\ over\ 856\ s=1.3\ MHz/1000\ s is close to that obtained in Section\ \ref{sec:DriftAnd2Scales} above, $d f/d t =1.5$\ MHz/1000\ s.
We will adopt the rate of 1.5 MHz/1000 s as more accurate  because the structure function for large time shift is weak and has a large variations. 

For the 
space-ground baseline (RA-GB), a time shift of the structure function produces only a decrease in the 
amplitude of the structure function, without displacement of its 
minimum, even with time shifts as large as 800 s. 
This is consistent with absence of the wide  structure on the long baseline.
We further conclude that the refraction that causes displacement of the structure function
takes place behind the screen responsible for diffractive scintillation.

\subsection{Spatial coherence function} \label{sec:SpatialCoherenceFunction}

According to Equation 10, $J_1(\vec b,\Delta f)$ is the second moment of $ j (\vec \rho, \vec \rho + \vec b,  f, t)$ and also is the fourth moment of the field $u(\vec \rho,  f, t)$.
As \citet{Pr75} showed, the fourth moment of the field can be expressed through the second moments and,  in the regime of strong scintillations 
\begin{align} 
J_1(\vec b,\Delta f) &= |\langle j (\vec \rho, \vec \rho + \vec b,  f, t) j^* (\vec \rho, \vec \rho + \vec b,  f + \Delta f, t) \rangle |  \nonumber \\ \quad &=   |B_u (f)|^2 + |B_u (\vec b)|^2.
\label{eq:16_J_1_def} 
\end{align} 
Here
$B_u (\vec b)$ is the spatial field-coherence function at a 
single average flux, and $B_u(f)$ is the frequency correlation function of 
fluctuations in flux, and is independent of baseline. If the spatial coordinate in the phase screen plane is $\Delta \vec \rho $ then \citep{Pr75}:
\begin{align} 
B_u (\Delta \vec \rho) = \exp[- {\textstyle{\frac{1}{2}}} D_s(\Delta \vec \rho)], 
\label{eq:15}  
\end{align} 
where
 $D_s (\Delta \vec \rho)$ is the spatial structure function of phase fluctuations:
\begin{align}
\label{eq:StructureFunctionDefinition}
D_s(\Delta \vec \rho)&=\langle \phi(\vec x +\Delta \vec \rho) -  \phi(\vec x ) \rangle_{\vec x}
\end{align}
where $\phi(\vec x)$ is the screen phase at $\vec x$. 
In the case of a spherical wave at the observer plane we have 
\begin{equation} 
D_s(\Delta \vec \rho) = \int \limits_{0}^{z} dz^\prime D \left(\Delta \vec \rho \right) 
\label{eq:16} 
\end{equation}
where D is a gradient of $D_s(\Delta \vec \rho)$ along the z-axis and 
$\Delta \vec \rho = \frac{(z-z^\prime ) }{ z}\vec b$. 
The integration is from the observer at $z^\prime = 0$ to the pulsar at $z^\prime = z$. 
 
For the space-ground interferometer we used Equation 15 with $b=b_s$, where $b_s$ is the interferometer baseline. According to Equation 15,
for $\Delta f=0$  
we have $J_1 (\vec b_s ,\Delta f = 0) = 1 + |B_u (\vec b_s)|^2$; and for $\Delta f > \Delta f_{dif}$ we have $J_1 (\vec b_s ,\Delta f > \Delta f_{dif}) = |B_u (\vec b_s)|^2$.
As displayed in Figure\ \ref{fig:fig3},
we find $J_1 (\vec b_s ,\Delta f =0) = 1.4\times 10^{-4}$
and $J_1 (\vec b_s ,\Delta f > \Delta f_{dif}) =  2.4\cdot 10^{-5}$ at $\Delta f$ = 2 MHz.
Thus,
\begin{align} 
\frac{J_1 (\vec b_s ,\Delta f > \Delta f_{dif})}{J_1 (\vec b_s ,\Delta f =0)} &=  \frac{\left|B_u (\vec b_s ) \right|^2}{1 + \left|B_u (\vec b_s ) \right|^2}   =  0.17
\label{eq:17} 
\end{align} 
From this equation, we obtain 
\begin{align} 
\left|B_u (\vec b_s ) \right|^2 & =  0.20 \label{eq:21}
% \\
% \left|B_u (\vec b_s ) \right| \; & =  0.45 \nonumber
\end{align} 

From the analogous calculation for the ground interferometer (Figure\ \ref{fig:fig6}a) we obtain 
\begin{equation} 
\frac{J_2 (\vec b_g,\Delta f > \Delta f_{dif} )}{J_2 (\vec b_g,\Delta f =0)}   =   \frac{|B_u (\vec b_g) |^2}{[1 + |B_u (\vec b_g) |^2]}   =  0.50
\label{eq:19} 
\end{equation} 
This implies 
\begin{equation} 
|B_u (\vec b_g ) |^2  = 1.
\label{eq:20} 
\end{equation} 
This result implies that the ground interferometer does not resolve the scattering disk; mathematically, it means that
$|b_g| \ll 1/k\theta_{scat}$.

\section{Model of the turbulent interstellar medium}\label{sec:Model}

Our analysis leads to the following model for the scintillation.  
The material responsible for scintillation of PSR\ B1919+21 consists of two components: 
strong diffractive scattering in a layer of turbulent 
density inhomogeneities at a distance of $z_1$, that is responsible for the small-scale structure in the spectra; 
and weak and refractive scattering in a  layer of turbulent density inhomogeneities close to the observer, at a distance of $z_2$, that is responsible for the large-scale 
structure. The spatial structure function of phase fluctuations 
$D_s(\Delta \rho)$, as described in Section\ \ref{sec:SpatialCoherenceFunction} above, characterizes the two screens.

\begin{figure}
\includegraphics[height=\columnwidth, angle=270]{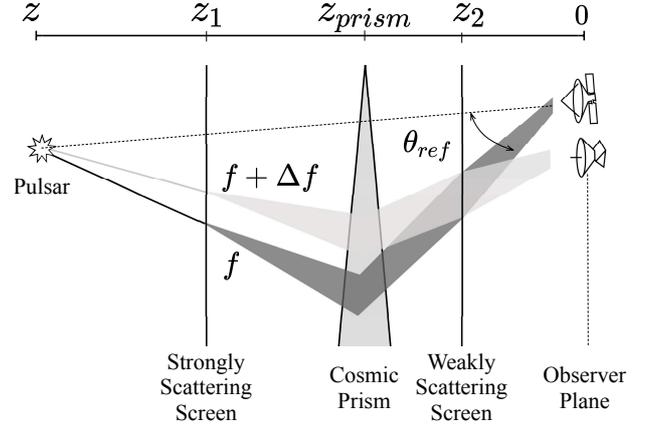}
\caption{Schematic illustration of model geometry, showing radiation at a lower frequency $f$ (dark line)
scattered by the strongly-scattered screen 1 into a range of angles (dark shading), deflected by the cosmic prism, and further scattered by the weakly-scattering screen 2 before reaching the observer plane.  Lighter path and shading shows a higher frequency $f+\Delta f$.}
\label{fig:geometry}
\end{figure}

We describe here a physical model for the distribution of scattering material that explains our observations.
Suppose we have a cosmic prism, located close to the observer at a 
distance of $z_{prism}$, which deflects the beam with an angle of 
refraction $\Delta \vec{\theta}_{ref}$. Let $\vec{\theta}_{ref,0}$ be the 
resulting shift of source position visible in the observer plane at 
frequency $\nu_0$. The difference in refraction angle at a nearby frequency $\nu_0 +  f$ is \cite{Sh03} 
\begin{equation} 
\Delta \vec \theta_{ref}  =  2 \frac{ f}{\nu_0} \vec{\theta}_{ref,0} \label{eq:25} 
\end{equation}

Suppose further that phase screen 2 is located close to the observer and the distance between the observer and the phase screen 2 is much smaller 
than  $z$, the distance between the observer and the pulsar:
\begin{equation} 
z_2 \ll z \label{eq:26} 
\end{equation} 
The cosmic prism also located close to the observer, but a little bit further that the phase screen 2 
\begin{equation} 
z_2 < z_{prism} \ll z \label{eq:27} 
\end{equation} 
Phase screen 1 is located much further away along the line of sight, at a distance of order $z/2$. 
The structure functions of phase 
fluctuations for the phase screens then have the model forms \citep{SSS98}:
\begin{align} 
D_{S,1}(\Delta \vec{\rho})  &=  (k \theta_{scat,1}|\Delta 
\vec{\rho}|)^{\alpha_1}, \label{eq:23} \\
D_{S,2}  (\Delta \vec{\rho})  &=   (k \theta_{scat,2}|\Delta 
\vec{\rho}|)^{\alpha_2}, \label{eq:24} 
\end{align} 
Thus, our model for the turbulent plasma towards the pulsar is characterized by the following parameters: 
$\vec{\theta}_{ref,0}, \theta_{scat,1}, \theta_{scat,2}, \alpha_1, 
\alpha_2, z_1 , z_2, z_{prism}$. 
Figure\ \ref{fig:geometry} illustrates the geometry.
 
This model has similar structure to that used for our 
studies of the scintillations of pulsar B0950+08 \cite{S14}. However, in 
the present case the distance to phase screen 1 is considerably 
greater, and the characteristic scattering angle $\theta_{scat,1}$ is
significantly larger. Thus, for pulsar B1919+21 the scintillations are strong 
and saturated, with modulation index close to 1, as discussed in Section\ \ref{sec:DriftAnd2Scales}. Accordingly, we apply
the theory of saturated scintillations \cite {Pr75}. 
% The average 
% value of $\langle j (\vec{\rho}, \vec{\rho} + \vec{b},  f, t) 
% \rangle$ 
In this case, the field coherence function $B_u(\vec b)$ is given by:
\begin{align} 
\left\langle j (\vec{\rho}, \vec{\rho} + \vec{b},  f, t) \right\rangle &=  
\left\langle u (\vec{\rho},  f, t ) u^* (\vec{\rho} + \vec{b},  f, t ) 
\right\rangle  \nonumber\\ &=  B_u ( \vec{b} ) \label{eq:28.1} 
\\
&=\exp\left[- {\textstyle{\frac{1}{2}}}  D_{S,1} \left({\textstyle{\frac{(z-z_1)}{z}}} \vec{b} \right)- {\textstyle{\frac{1}{2}}} D_{S,2} ( \vec{b}) \right] 
\nonumber
\end{align} 
For the distant screen 1, the sphericity of the wavefront at the screen is important, and the
conversion of the baseline in the observer's plane $\vec {b}$ to the distance between the beams in the phase screen plane 
 is given by the equation: 
\begin{equation} 
\Delta \vec{\rho}_{1,b}    =   \frac{(z - z_1)}{z} \vec{b} 
\label{eq:29.1} 
\end{equation} 
 For the closer phase screen 2 we can neglect the sphericity 
factor. 

For saturated scintillations, the second moment $j (\vec{\rho}, \vec{\rho} + \vec{b},  f, t) $ consists of two 
 components: diffractive and refractive \citep{Pr75}. 
The diffractive component can be represented as 
\begin{align} 
J(\vec{b}, \Delta f, \Delta t)_{dif}  & =  \langle j (\vec{\rho}, \vec{\rho} 
+ \vec{b},  f, t) \nonumber\\ & \quad \times j^*(\vec{\rho}, \vec{\rho} + \vec{b},  f + \Delta f, t + 
\Delta t) \rangle _{dif} \label{eq:28.2} 
 \\ 
 & =  \langle j (\vec{\rho}, \vec{\rho} + \vec{b},  f, t) \rangle 
\langle j^*(\vec{\rho}, \vec{\rho} + \vec{b},  f, t) \rangle  \nonumber \\ &\quad + B_j (\Delta f, \Delta t) \nonumber
\end{align} 
 Here $ B_j (\Delta f, \Delta t)$ is a frequency-time correlation function of flux fluctuations independent of the projected baseline:
\begin{align} 
B_j (\Delta f, \Delta t) &=  \langle u(\vec{\rho},  f, t) 
u^*(\vec{\rho},  f  + \Delta f, t +\Delta t) \rangle  \nonumber\\ 
& \ \, \times  \langle u 
(\vec{\rho} + \vec{b},  f, t)   u^*( \vec{\rho} + \vec{b},  f + \Delta f, t 
+\Delta t) \rangle   \label{eq:29.2}\\
& = |B_u(\Delta f, \Delta t)|^{2} 
 \nonumber
\end{align} 

Pulsar motion 
at a transverse speed of $\vec{V_p}$
is primarily responsible for 
variations of flux density with time, producing scintillations from phase screen 1 at distance 
$z_1$. Pulsar motion leads to a shift of the beam in the plane of the phase screen:
\begin{equation} 
\Delta \vec{\rho}_{1,t}  =\frac{z_1}{z} \vec{V}_p \Delta t 
\label{eq:30} 
\end{equation} 
For 
$\Delta f = 0$ we have 
\begin{align} 
B_u (\Delta t) 
&= \exp\left[-  {\textstyle{\frac{1}{2}}} D_{S,1}(\Delta \vec{\rho}_{1,t})\right] \nonumber \\&= \exp\left[-{\textstyle{\frac{1}{2}}}\left( \frac{\Delta t}{\Delta t_{dif}} \right)^{\alpha_1}\right] \label{eq:33} 
\end{align} 
with
\begin{equation} 
\Delta t_{dif}  =  \frac{z}{z_1 k\, \theta_{scat,1} |\vec{V_p}|} \label{eq:32a} 
\end{equation} 
Correspondingly, 
\begin{equation} 
B_j (\Delta t)  =  \exp[-  D_{S,1}(\Delta \vec{\rho}_{1,t})] =  \exp[- (\Delta t/\Delta t_{dif} )^{\alpha_1}] \label{eq:33a} 
\end{equation} 
 
For a ground baseline, the projected length $b_g$ is much 
smaller that the coherence scale of the field ${1}/{k \theta_{scat,1}}$:
\begin{align} 
b_g  \ll  \frac{1}{k \theta_{scat,1}} \label{eq:36} 
\end{align} 
and correspondingly, 
\begin{equation} 
B_u (\vec{\rho}, \vec{\rho} + \vec{b_g}) = 1 .
\label{eq:37} 
\end{equation} 
For a space-ground baseline, the projected length $b_s$ is larger than
the coherence scale of the field:
\begin{equation} 
b_s > \frac{1}{k \theta_{scat,1}}\label{eq:38} 
\end{equation} 
and correspondingly, 
\begin{equation} 
B_u (\vec{\rho}, \vec{\rho} + \vec{b_s}) = 
\exp \left[ - \frac{1}{2} \left( \frac{z - z_1}{z} \cdot  
\frac{b_s}{k \theta_{scat,1}}\right)^{\alpha_1}\right]  \label{eq:37a} 
\end{equation} 
Here, $[(z-z_1)/z]$ is the sphericity factor, which converts 
the baseline $b_s$ into the distance between beams in the phase 
screen plane at distance $z_1$. Comparison of $J (\vec {b}, \Delta f = 0)$ and $J(\vec b,\Delta f > \Delta f_{dif})$ 
allows us to estimate the spatial coherence function (interferometer visibility), as in Section\ \ref{sec:SpatialCoherenceFunction} above. 

\section{Results} 
 
As mentioned above, the time correlation function of 
interferometer response fluctuations is determined primarily by pulsar 
motion with transverse velocity $\vec{V}_p$. Pulsar motion shifts the beam 
in the plane of phase screen 1 by $\Delta 
\vec{\rho}_{1,t}$ (Equation \ref{eq:30}). 
The temporal correlation function of flux density shown in Figure\ \ref{fig:fig4} yields 
$\Delta t_{dif} = 290\ {\rm s}$. 
We fit the model given by Equation\ \ref{eq:33a} to the shift of the structure function shown in Figure 7  to find the index $\alpha_1$. The fit yielded $\alpha_1 = 1.73$. 
The normalized spatial correlation 
function of flux fluctuations for the space-ground baseline $\vec b_s$ is (Equation\ \ref{eq:21}) :
$|B_u(\vec {b_s}|^{2}$  =  0.20. 

From the projected length of the ground-space baseline $b_s = 6 \times 10^9$ cm we find 
$b_{dif}$:
\begin{equation} 
b_{dif} = \frac{z}{(z-z_1)} \frac{1}{k \theta_{scat,1}} = 4.6\times 10^9\ {\rm cm} 
\end{equation} 
Using the measured proper motion of $\mu_{\alpha}$ = 17 $\pm 4\ {\rm mas\ yr}^{-1}$, 
$\mu_{\delta}$ = 32 $\pm 6\ {\rm mas\ yr}^{-1}$ \citep{Z05}, and an assumed distance to the 
pulsar $z = 1 {\rm\ kpc}$ \citep{CL}, we obtain a pulsar tangential velocity of
$\vec{V}_p$ = 200 km/s. Comparison of $\left| \vec{V}_p\right|\, \Delta t_{dif}$ with
$b_{dif}$ yields $z_1/(z-z_1) = b_{dif}/\left| \vec{V}_p\right|$ = 0.78. 
Hence, $z_1 = 0.44\,z$ = 440 pc. Therefore, the screen is located 
approximately halfway between the pulsar and the observer. From knowledge of  
$b_{dif}$ and $z_1$, we obtain $\theta_{scat,1} = 1.2 {\rm\ mas}$. In 
the observer plane, $\theta_{obs,1} = [(z-z_1)/z]\theta_{scat,1} = 0.7 {\rm\ mas}$. 

The normalized frequency correlation function $R(\Delta f)$ of intensity variations is 
determined by the diffractive scintillations. 
\cite {O77} found that $R (\Delta f)$ is: 
\begin{align} 
R(\Delta f) &= 1 -{\textstyle{\frac{1}{2}}} \left(\frac{\Delta f}{\Delta f_{dif}}\right)^{\alpha_1/2}, \quad \Delta f < 2 \Delta f_{dif}  \\
\Delta f_{dif} &= \frac{c}{4\pi 
A(\alpha_1)\, z\, (z_1/(z-z_1))\, (\theta_{obs,1})^2} \nonumber
\end{align} 
Taking $\alpha_1$ = 1.73, we find for the constant $A$:
\begin{equation} 
A(\alpha_1) = [2 \Gamma(1+\alpha_1 /2)cos(\pi\alpha_1 
/4)]^{\frac{2}{\alpha_1}} \approx 0.34 
\end{equation} 
where $\Gamma$ is the complete gamma function. Using our estimated values for $z_1$ and
$\theta_{scat,1}$, we find $\Delta f_{dif} = 290 {\rm\ kHz}$, which 
coincides very well with our measured value of
$\Delta f_{dif}$ = 330 kHz. 

If we suppose the scattering material is homogeneously
distributed between observer and pulsar, then the 
frequency diffraction scale will be determined by the relations:
\begin{align} 
\Delta f_{dif} &= \frac{c}{4\pi B(\alpha_1)z(\theta_{obs,1})^2}  \\
B(\alpha_1) &= \left[\frac{ 2 (\Gamma(1+\alpha_1 /2))^3 \cos(\pi \alpha_1 /4)(1+\alpha_1) 
}{\Gamma(2 + \alpha_1)}\right]^{ {2}/{\alpha_1}}  \nonumber\\ &\approx 0.18   
\end{align} 
We obtain $\Delta f_{dif}$ = 430 kHz, which also coincides with 
the observations. 
Thus, our measurement of $\Delta f_{dif}$ agrees with either a thin-screen model 
(located at $z_1 \approx 0.44 z$) 
or with  
homogeneously distributed scattering material, to about the $30\%$ accuracy of our measurement.
The consistency of the measured and calculated frequency scales suggests that the 
assumed pulsar distance of $z=1$\ kpc corresponds to the actual distance. 
 
The spatial correlation function for weak scintillations from the  
inhomogeneities of the layer $z_2$ can be represented as \citep{S14} 
\begin{align} 
R(\Delta \vec{\rho}_2) &= D_{S,2}(\Delta \rho_{2,Fr}) - 
D_{S,2}(\Delta \vec{\rho}_2) \\
& = D_{S,2}(\Delta \rho_{2,Fr})\left[1-\left(\frac {\Delta 
\rho_2}{\Delta \rho_{2,Fr}}\right)^{\alpha_2}\right], \nonumber \\ & \quad {\rm with:\ }  |\Delta \vec{\rho}_2| \ll \Delta
\rho_{2,Fr}  \nonumber
\end{align}
where:
\begin{align}
\Delta \rho_{2,Fr} &= \sqrt{\frac{z_2}{k}} 
\end{align} 
Here $\Delta \vec{\rho}_2$ is the separation of points in the 
observer's plane (and, equivalently, in the near phase screen at distance $z_2$). 

The frequency-time 
correlation function can be obtained by the replacement $\Delta 
\vec{\rho}_2 \rightarrow \Delta \vec{\rho}_{2,t} + \Delta 
\vec{\rho}_{2,f}$, where 
\begin{align} 
\Delta \vec{\rho}_{2,\Delta f} &= -2z_2({\Delta f}/{\nu_0})\vec \theta_{ref,0}\\ 
\Delta \vec{\rho}_{2,\Delta t} &= \vec{V}_{obs}\Delta t
\end{align}
Here $\vec{V}_{obs}$ is the observer's velocity.  Figure \ref{fig:fig1} shows that 
the frequency structure of the diffraction pattern drifts in time, with the 
speed ${df}/{dt}$ = 1.5 MHz/1000 s, so that diffraction spots are extended correspondingly in the dynamic spectrum.
 The component of velocity $\vec V_{obs}$ parallel to the refraction angle $\vec \theta_{ref,0}$ defines this drift.
 It produces the shift with time lag of the minimum in frequency of the structure function, as shown in Figure\ \ref{fig:ground_space_freq_deltat_SF}. 
 The component of velocity perpendicular to $\vec \theta_{ref,0}$ does not contribute to the shift, but does produce an asymmetry of the structure function about the frequency lag of the minimum, $f_{min}$.
Specifically, the structure function becomes flatter for frequency lags smaller than $f_{min}$ and steeper for frequency lags more than $f_{min}$. 
It is difficult to distinguish such an asymmetry at large time shifts because of strong influence of noise. However when 
$\vec V_{obs}$ is parallel to the refraction angle $\vec \theta_{ref,0}$ features in the dynamic spectra are strongly elongated, 
as we observe.  This suggests that the perpendicular component of velocity is small compared with the parallel component, and we conclude that
 the vectors $\vec{V}_{obs}$ and 
$\vec{\theta}_{ref,0}$ are approximately parallel. In 
this case, we can represent the frequency-time correlation function as 
\begin{equation} 
R(\Delta \vec{\rho}_2) = D_{S,2}(\Delta \vec{\rho}_{2,Fr})\left( {1 - 
{\textstyle{\frac{1}{2}}} \left[\frac{t}{t_{2,0}} - \frac{f}{f_{2,0}}\right]^{\alpha_2}}  \right)
\label{eq:48}
\end{equation} 
From the shift of $f_{min}$ for the largest time shift of $\Delta t =640\ P_1$ (Figure 9), 
$f_{2,0} = 1.1 {\rm\ MHz}$, we find $t_{2,0} = 
f_{2,0}({df}/{dt})^{-1} = 700{\rm\ s}$. 

The observer's velocity relative to the Local Standard of Rest,
perpendicular to the pulsar's line of sight
for the date of our 
observation, was $\vec{V}_{obs} = 23.7 {\rm\ km\ s}^{-1}$. Using this value, we 
find for the Fresnel scale $\Delta \rho_{2,Fr} = 2^{1/{\alpha_2}} 
V_{obs}t_{2,0} = 2.5\times 10^9{\rm\ cm}$, using $\alpha_2 = 1.73$. 
The observer velocity is the vector sum of the velocity of orbital motion of the Earth on the date of observation and the
 velocity of the Sun relative to the Local Standard of Rest projected to perpendicular to the pulsar's line of sight. 
We do not know the velocity of the
 clouds of turbulent plasma responsible for scintillation, but we assume they have velocity relative to the observer of $|\vec V_{scr}|\approx 10{\rm \ km\ s}^{-1}$ or less,
although we do not know its direction and magnitude. The Fresnel scale we find corresponds to
 $\vec V_{scr}$ = 0. If we assume that the screen has the velocity of 10 km/s the error in evaluation of  $\Delta \rho_{2,Fr}$  will be 0.75$\times 10^9$ cm (about 30\%). 
Accordingly, the distance to the near layer is $z_2$ = 0.14 $\pm 0.05$ pc. Using 
 Equations 47 and 49 with $\Delta f =  f_{2,0}$, we obtain $\theta_{ref, 0} = 110 \pm 30$ \rm mas. 

We found above that the relative 
amplitude of the second component is 0.15. 
Hence, 
\begin{equation} 
D_{S,2}(\Delta \rho_{2,Fr}) = (k\theta_{scat,2}\Delta 
\rho_{2,Fr})^{\alpha_2} = 0.15 
\end{equation} 
For $\alpha_2$ = 1.73 we find $\theta_{scat,2} = 0.4{\rm\ mas}$. The error of $\theta_{scat,2}$ is about 30\%. 
 
From the condition that the cosmic prism
 has no significant effect 
on the frequency correlation of diffractive scintillations, we can 
estimate an upper limit for the distance to the cosmic prism $z_ 
{prism}$. 
Change in the refraction angle of the cosmic prism with change
of frequency $f$ displaces the 
diffraction pattern from the strongly-scattering far screen.  
We assume that the frequency scale of the pattern $\Delta f_{dif} $ is less than 
the offset of the scattering from refraction. We then obtain the 
displacement 
\begin{equation} 
2(\Delta f_{dif} / \nu_0) z_{prism} \theta_{ref, 0} < b_{dif} = 
4.6 \times 10 ^ 9 {\rm \ cm} 
\end{equation} 
Substituting our observed values of 
$\Delta f_{dif}  = 330 {\rm\ kHz}$ and $ \theta_ {ref, 0}  = 
110 {\rm\ mas}$ into this inequality, we find $ z_ {prism} \le $ 1.4 pc, or $ z_ {prism} \le $ 2 pc when we include the error in our estimate of $\theta_{ref}$. 

Thus, 
we find 3 components that contribute to scintillation of pulsar B1919+21:
distant material at $z_1 \approx 440$\ pc, or material homogeneously distributed along the line of sight
to the pulsar;
a cosmic prism at a distance $z \le  2$\ pc, and a nearby screen at $z_2=0.14$\ pc.
Most pulsars show some scattering distributed along the line of sight, and the distant material indicates that B1919+21 is no exception.

Cosmic prisms are seen for a number of nearby pulsars.
\citet{Sh03} found the first evidence for such a component, from 
analysis of multifrequency observations they found a refraction angle in the direction to PSR B0329+54 of about 0.6 mas at frequency 1 GHz. 
They inferred that the size of
irregularities responsible for refraction is less or about 3$\times 10^{15}$ cm. 
\citet{SG06} found an
indication that refractive effects dominate scattering for the direction to PSR J0437$-$4715. 
\citet{S14} found a cosmic prism in the direction of PSR B0950+08 using 
space-ground interferometry.
The refraction angle was measured as 1.1 to 4.4\ mas at frequency 324 MHz. 
Here, we report the first localization of a cosmic prism, at a distance of about 1.4\ pc
in the direction to PSR B1919+21.
The material associated with this prism is unknown.
However, distances of only a few pc are inferred for the material that is responsible for the scintillation of intra-day variable extragalactic sources 
\citep{KC, D02, big03, den03, Jau03, B06}.
This material may lie at interfaces where nearby molecular clouds collide \citep{LR}.

The distance that we find for the close screen, of only 0.14\ pc, is extremely close.
It lies hundreds of times further away than the termination shock of the solar wind, but within the Oort cloud, and hence within our Solar System.
Our observation is the first detection of scattering by ionized gas in this region.
Clearly, additional observations are needed to clarify the position and distribution of this material, and its relation to other plasma components of the Solar System and the solar neighborhood.

\section{Conclusion} 

We have successfully conducted  space-ground observations of PSR 
B1919+21 at frequency 324 MHz with a projected space-ground baseline of 
60,000 km. 
Analysis of frequency and time correlation functions and structure functions provides
an estimate of the spatial distribution of 
interstellar plasma along the line of sight. We show that the 
observations indicate the existence of two
components of scattering material in this direction. One is a screen located at a distance of 
about 440 pc from the observer, or distributed homogeneously along the line of sight.
This shows strong diffractive scintillations and produces the largest effect.
 The second component is in a much closer
screen, at a distance of about 0.14 pc, \rm and 
corresponds to weak scintillation. The Fresnel scale is equal to 
$2.5\times 10^9 {\rm\ cm}$ at the near screen. Furthermore, a cosmic prism is located beyond the 
near screen, leading to a drift of the diffraction pattern across the dynamic spectrum at
a speed of $df/dt = 1.5 {\rm\ MHz}/1000{\rm\ s}$. 
We have estimated the refraction angle of this prism
as $\theta_{ref,0} = 110{\rm\ mas}$, and obtained an upper limit for the 
distance to the prism: $z_{prism} \le  2 {\rm\ pc}$. \rm  Analysis of the spatial 
coherence function for the space-ground baseline (RA-GB) allowed us to 
estimate the scattering angle in the observer plane: 
$\theta_{scat} = 0.7{\rm\ mas}$. From temporal  and frequency structure 
functions analysis we find for the index of interstellar plasma 
electron density fluctuations to be $n = 3.73.$ 

\section*{Acknowledgements}

The RadioAstron project is led by the Astro Space Center of the 
Lebedev Physical Institute of the Russian Academy of Sciences and 
the Lavochkin Scientific and Production Association under a 
contract with the Russian Federal Space Agency, in collaboration 
with partner organizations in Russia and other countries. We are 
very grateful to the staff at the Westerbork synthesis array and Green Bank observatory \rm for 
their support. The study was supported by the program of the Russian 
Academy of sciences 'Nonsteady and explosive processes in Astrophysics'.
C.R.G.\ acknowledges support of the US National Science Foundation (AST-1008865).
 
 {\it Facilities:} {RadioAstron Space Radio Telescope (Spektr-R), GB, WB}.


%\bibliographystyle{mnras}
%\bibliography{1919} % if your bibtex file is called example.bib

%%%%%%%%%%%%%%%%%%%%%%%%%%%%%%%%%%%%%%%%%%%%%%%%%%

%%%%%%%%%%%%%%%%% APPENDICES %%%%%%%%%%%%%%%%%%%%%

%\appendix



%%%%%%%%%%%%%%%%%%%%%%%%%%%%%%%%%%%%%%%%%%%%%%%%%%

% Don't change these lines
\bsp	% typesetting comment
\label{lastpage}
\end{document}